# Slippery Polymer Monoliths: Surface Functionalization with Ordered MoS$_2$ Microparticle Arrays

Weijia Han[a,*], Siwei Luo[b,d], Dirk Bröker[c], Norbert Vennemann[c], Markus Haase[a], Georg S. Duesberg[b], and Martin Steinhart[a,*]

[a] *Institut für Chemie neuer Materialien and CellNanOs, Universität Osnabrück, Barbarastr. 7, 49076 Osnabrück, Germany*

[b] *Institute of Physics, EIT 2, Faculty of Electrical Engineering and Information Technology Universität der Bundeswehr München Werner-Heisenberg-Weg 39, 85577 Neubiberg, Germany*

[c] *Fakultät Ingenieurwissenschaften und Informatik, Hochschule Osnabrück, Albrechtstr. 30, 49076 Osnabrück, Germany*

[d] *School of Physics and Optoelectronics, Hunan Key Laboratory for Micro-Nano Energy Materials and Devices, Laboratory for Quantum Engineering and Micro-Nano Energy Technology, Xiangtan University, Hunan 411105, People's Republic of China*

\* Corresponding author.

  *E-mail addresses:* weijia.han@uni-osnabrueck.de (W. Han), martin.steinhart@uni-osnabrueck.de (M. Steinhart).

# ABSTRACT

Components of technical systems and devices often require self-lubricating properties, which are implemented by means of dry lubricants. However, continuous lubricant coatings on the components' surfaces often suffer from poor adhesion, delamination and crack propagation. The replacement of continuous coatings with dense ordered arrays of microparticles consisting of dry lubricants may overcome these drawbacks. Using the well-established solid lubricant $MoS_2$ as model system, we demonstrate that the sliding capability of polymeric monoliths can be significantly enhanced by integration of arrays of micron-sized dry lubricant microparticles into their contact surfaces. To synthesize the $MoS_2$ microparticle arrays, we first prepared ordered hexagonal arrays of ammonium tetrathiomolybdate (ATM) microparticles on Si wafers by molding against poly(dimethylsiloxane) templates followed by high-temperature conversion of the ATM microparticles into $MoS_2$ microparticles under $Ar/H_2$ atmosphere in the presence of elemental sulfur. Finally, the obtained large-scale hexagonal $MoS_2$ microparticle arrays were transferred to the surfaces of polymer monoliths under conservation of the array ordering. Self-lubrication of components of technical systems by incorporation of dry lubricant microparticle arrays into their contact surfaces is an example for overcoming the drawbacks of continuous functional coatings by replacing them with microparticle arrays.

# 1. Introduction

The control of surficial friction of displaceable components in mechanical systems is a ubiquitous technical problem apparent in microscale electromechanical building blocks as well as in large-scale technical systems. Self-lubricating hybrid and composite materials capable of sliding against counter-bodies in the absence of liquid lubricants have attracted significant interest [1, 2]. Commonly, self-lubrication is achieved with dry lubricants, which are either deposited on or incorporated into the displaceable components. Self-lubricating components often consist of hybrid materials containing dry lubricants within their entire volumes. Then, the material properties of the other components may deviate from the bulk material properties in an undesired way. Moreover, it might be challenging to concentrate sufficient amounts of the dry lubricants at the sliding surfaces. Finally, the unnecessary presence of solid lubricant away from the sliding surface may result in higher material costs. It is, therefore, desirable to locate dry lubricants exclusively at the sliding surfaces. However, film-like dry lubricant coatings prepared by methods such as physical and chemical vapor deposition may suffer from poor adhesion to their substrates.[3-5] Hence, buckling, wrinkling, delamination and crack propagation may occur. These drawbacks may be circumvented by incorporating solid lubricants in a particulate form selectively into the sliding surface. To ensure uniform friction properties, the lubricant particles should be uniform in size and evenly distributed across the sliding surface.

One of the most important dry lubricants is molybdenum disulfide ($MoS_2$) [6-8], a chemically inert transition metal dichalcogenide that has also been employed in catalysis [9-11], electronics [12, 13], photonics and optoelectronics [14, 15], bio-imaging and biosensing [16, 17], as well as in medical applications [16, 18]. $MoS_2$ consists of stacks of two-dimensional $MoS_2$ layers held together by van der Waals interactions. Because of easy interlayer sliding, $MoS_2$ possesses low friction coefficients as well as low shear strengths. As a lubricant, $MoS_2$ is vacuum-resistant and chemically stable at least up to 350 °C in oxidizing environments [19]. It would, therefore, be highly attractive to incorporate dense and regular arrays of $MoS_2$ particles into the sliding surfaces of bulk parts and components. Synthetic methods for the preparation of $MoS_2$ nanoparticles have been summarized in recent reviews [20-24], some of which put special emphasis on the use of $MoS_2$ nanoparticles as lubricant additives [25, 26]. Few activities have been directed to the rational positioning of $MoS_2$ nanostructures and microstructures, which are often two-dimensional rather than three-dimensional, onto substrates. Thin elemental Mo layers patterned *via* top-down lithography were converted to MoS$_2$ by heat treatment in the presence of H$_2$S [27]. Arrays of 2D MoS$_2$ flakes were prepared by lithographic patterning combined with dry etching of MoS$_2$ sheets [28] or of chemical conversion of MoS$_2$ precursors [29].

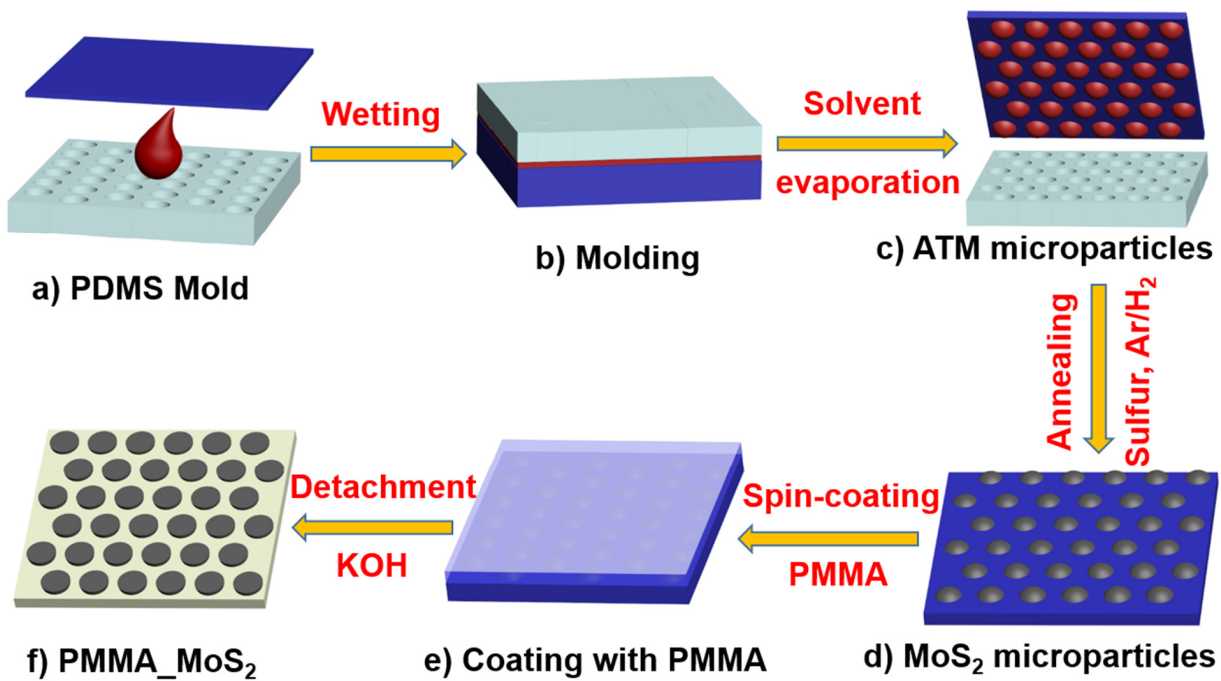

**Fig 1**. Schematic diagram of the preparation of PMMA monoliths having a contact surface functionalized with arrays of embedded MoS$_2$ microparticles (PMMA_MoS$_2$ monoliths). a) A solution of ATM in DMF (red) is deposited onto a hydrophilized macroporous PDMS mold (light blue), which is then b) covered by a hydrophilized SiO$_2$-covered Si waver piece (blue). c) After evaporation of the DMF and detachment of the hydrophilized macroporous PMDS mold, a regular array of ATM microparticles (red) remains on the surface of the SiO$_2$-covered Si wafer piece. d) Thermal sulfurization converts the ATM microparticles into MoS$_2$ microparticles (gray), whereby the array ordering is conserved. e) The MoS$_2$ microparticles are covered with PMMA by spin-coating a PMMA/chloroform solution (white-transparent) onto the MoS$_2$ microparticle array. f) After detachment from the hydrophilized SiO$_2$-covered Si waver piece, a PMMA_MoS$_2$ monolith is obtained.

Here, we report the incorporation of dense, regular MoS$_2$ microparticle arrays into the surfaces of polymer monoliths to optimize their sliding properties (Fig. 1). We first prepared monodomains of ammonium tetrathiomolybdate (ATM) microparticles on Si wafers. Then we converted the ATM microparticles into crystalline MoS$_2$ microparticles under conservation of the array order by adapting a previously reported high-temperature reaction of ATM in the presence of elemental sulfur [30]. The ordered $MoS_2$ microparticle arrays obtained in this way were embedded onto the surface of poly(methyl methacrylate) (PMMA) monoliths (thereafter referred to PMMA_$MoS_2$ monoliths). Thus, under a preset contact pressure of 3858 Pa the friction coefficient was halved as compared to unmodified PMMA monoliths.

## 2. Materials and Methods

*2.1. Materials*

Macroporous silicon [31, 32] with conical macropores having hemispherical macropore bottoms (base diameter: 960 nm; bottom diameter: 650 nm; macropore depth: 1.0 µm; interpore distance: 1.5 µm; cf. Fig. S1) was purchased from SmartMembranes (Halle, Saale). (100)-oriented *p*-type silicon wafers (1-30 Ωcm) coated with 300 nm thick dry $SiO_2$ layers were purchased from Siegert Wafer GmbH (Aachen, Germany). ATM (($NH_4)_2MoS_4$; 99.97 %), polystyrene (PS; $M_n$ = 267.000 g/mol; $M_w$ = 253.000 g/mol; $M_w/M_n$ = 1.05), KOH (85.5 %), 1H,1H,2H,2H-perfluorodecyltrichlorosilane (PFDTS; 97 %), sulfur (99.98 %), chloroform (99 %), and *N,N*-dimethylformamide (DMF, 99.8 %) were purchased from Sigma-Aldrich. PMMA ($M_w$ = 1.820.000 g/mol; $M_n$ = 1.580.000 g/mol; $M_w/M_n$ = 1.15) was supplied by Polymer Standards Service (Mainz, Germany).

*2.2. Fabrication of macroporous PDMS molds*

Macroporous silicon pieces with an area of 2 × 2 $cm^2$ were cleaned by immersion into piranha solution (98 % $H_2SO_4$ and 30 % $H_2O_2$ at a volume ratio of 3:1) at 80 °C for 30 min and then copiously rinsed with deionized water. The surface of the macroporous silicon was modified with PFDTS by vapor deposition in a sealed container at 85°C for 2 h and subsequently at 130 °C for 3 h. Then, 80 µL of a 5 wt-% solution of PS in chloroform was poured onto PFDTS-modified macroporous silicon. The chloroform was allowed to evaporate overnight. The obtained negative PS replicas (Fig. S2) of the macroporous silicon were nondestructively detached and then covered with a precursor mixture for elastomeric PDMS (SYLGARD 184 formulation obtained from Dow Corning, USA; base and curing agent were mixed at a weight ratio of 10:1). After aging under ambient conditions for two days and nondestructive detachment, we obtained macroporous PDMS molds with hexagonal arrays of hemispherical macropores that were positive replicas of the macroporous silicon (Fig. S3).

*2.3. Preparation of ATM microparticle arrays*

$SiO_2$-covered Si substrates extending 2 × 2 $cm^2$ were sonicated in acetone, ethanol as well as deionized water for 10 min and dried in a nitrogen flow. The Si substrates and as well as the macroporous PDMS molds were hydrophilized by oxygen plasma treatment at 100 W for 4 min using a Diener femto plasma cleaner. Then, 1 μL of a 6.2 mM $(NH_4)_2MoS_4$ solution in DMF was deposited onto hydrophilized macroporous PDMS molds and covered with $SiO_2$-covered Si substrates treated as described above. After drying for 1 h under ambient conditions, the hydrophilized macroporous PDMS molds were manually detached and cleaned for reuse by sonification in DMF, ethanol and water.

*2.4. Synthesis of $MoS_2$ microparticle arrays*

$SiO_2$-covered Si substrates patterned with ATM microparticle arrays were placed in the central heating zone of a tube furnace in the presence of 1.5 g elemental sulfur at the upstream end. The furnace was heated to 700 °C at a rate of 30 K $min^{-1}$ under flow of a mixture of argon (100 s.c.c.m) and hydrogen (25 s.c.c.m). Under the same conditions, the samples were annealed at 700 °C for 60 min and then cooled to room temperature at the natural cooling rate of the tube furnace.

*2.5. Surface functionalization of PMMA monoliths with $MoS_2$ microparticle arrays*

We spin-coated 20 μL of a 5 wt% PMMA solution in chloroform onto the $MoS_2$ microparticle arrays located on $SiO_2$-covered Si substrates at 500 rpm for 5 s and then at 3000 rpm for 10 s. Subsequently, the samples were placed in an oven preheated to 180 °C and kept at this temperature for 10 min. After cooling to room temperature, the $SiO_2$ layers were etched in aqueous 30 wt-% KOH solution for 20 min. The PMMA_$MoS_2$ monoliths were floated on the solution surface and rinsed with deionized water for several times. As reference samples, pure PMMA monoliths were prepared following the same procedure using $SiO_2$-coated Si pieces without $MoS_2$ microparticle arrays.

*2.6. Measurements of friction coefficients of PMMA_$MoS_2$ monoliths*

Circular PMMA_MoS$_2$ monoliths with a diameter of 1.8 cm and a thickness of 115 μm (Fig. S4) and pure PMMA monoliths with the same dimensions were investigated with an Advanced Rheometric Expansion System (ARES) manufactured by T.A. Instruments in the constant strain mode at 24 °C. The samples were glued onto the center of the lower disk (diameter: 2 cm) of a motor shaft using UHU 48720-Drying Super Glue-Supergel. The upper aluminum disk (diameter: 2 cm) connected to the transducer shaft was slowly moved down until – after contact formation with the entire surface of the sample – preset normal forces were reached. Measurements were carried out at normal forces of 0.49 N, 0.98 N and 1.47 N corresponding to contact pressures of 1929 Pa, 3858 Pa and 5787 Pa. The lower plate was rotated with an angular frequency of 2.5π rad s$^{−1}$. The torque of PMMA_MoS$_2$ and PMMA monoliths prepared in the same way but in the absence of MoS$_2$ microparticles was measured and recorded using the software Orchestrator. Each measurement with a duration of 17 s was repeated for seven times.

*2.7. Morphological and structural characterization*

For scanning electron microscopy (SEM) a Zeiss AURIGA microscope was used and operated at an accelerating voltage of 7 kV. The geometrical descriptors apparent area, aspect ratio, roundness, and nearest neighbor distances of ATM microparticles (Fig. 2b) and MoS$_2$ microparticles (Fig. S5) imaged by SEM were determined with the software ImageJ using the freehand selection tool. The SEM Zeiss AURIGA was equipped with an INCA 350 energy-dispersive X-ray (EDX) spectroscopy unit (Oxford Instruments). Topographical images were acquired with an atomic force microscope (AFM) NT-MDT NTEGRA in the semi-contact mode using GOLDEN SILICON cantilevers (NSG01/Pt; NanoLaboratory) with a tip height of 14-16 μm, a tip curvature radius of 35 nm, a 20-30 nm thick reflective Pt side coating, and a resonant frequency of 87 kHz - 230 kHz. For X-ray powder diffractometry (PXRD) measurements, the MoS$_2$ microparticles were scratched from the SiO$_2$-covered Si substrates and collected as powder. The PXRD measurements were conducted on an X'Pert Pro diffractometer (PANalytical) with Bragg−Brentano geometry using Cu Kα (λ= 1.5406 Å) radiation (40 kV, 40 mA) and a 2θ step size of 0.0334°. Raman microscopy of MoS$_2$ microparticle arrays was carried out with a WITec alpha300 R confocal Raman imaging system (30 cm focal length; 600 grooves per millimeter grating spectrometer; spot diameter ~ 3 μm) equipped with an electron-multiplying charge-coupled device (EM-CCD; Andor Newton DU970N-BV-353) using the 532 nm line of a He–Ne laser with a power of 1.64 mW and an integration time of 0.5 s. For mappings, Raman spectra with an integration time of 0.2 s were acquired. The Raman spectra were analyzed and processed with the software Project FIVE. The stitched image of a MoS$_2$ microparticle array on a SiO$_2$-covered Si wafer shown in Fig. S6 was taken with an objective of Zeiss EC Epiplan-Neofluar 100x/0.9 using the built-in ocular video camera and the stitching function of the system. The stitching microscopy image (1000 μm × 1000 μm) composed of 361 individual high-resolution images was recorded by the software Project FIVE.

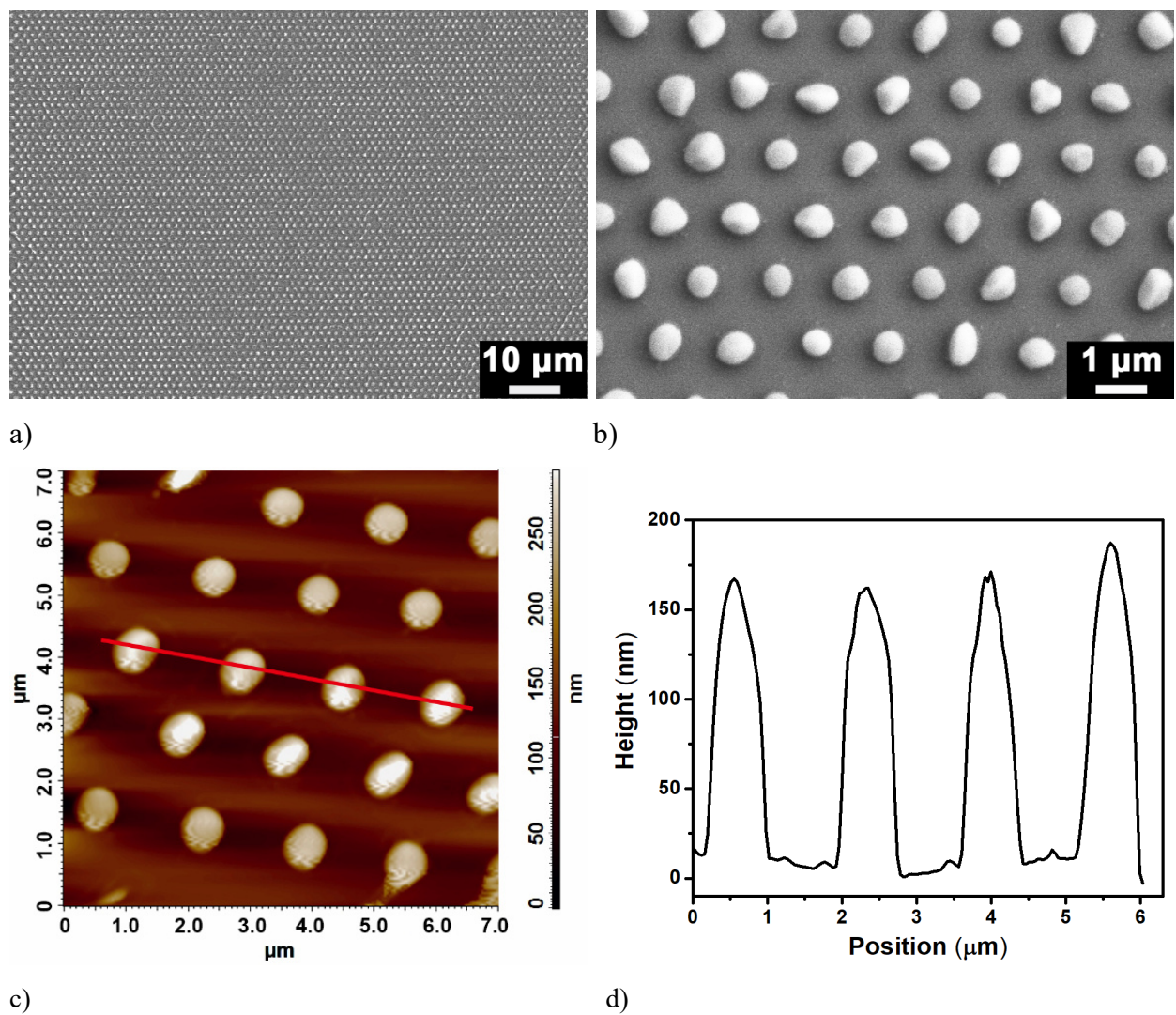

**Fig. 2.** ATM microparticle arrays on SiO$_2$-covered Si wafers. a) Large-area scanning electron microscopy (SEM) image; b) higher-magnification SEM image; d) Atomic force microscopy (AFM) topography image; d) topographic profile along the red line in panel c).

## 3. Results and discussion

The preparation algorithm for PMMA_MoS$_2$ monoliths is displayed in Fig. 1. At first, macroporous PDMS molds were prepared by two-step replication molding of macroporous silicon (Fig. S1) [31, 32] and hydrophilized by exposure to oxygen plasma. The macroporous PDMS molds were topographically patterned with hexagonal arrays of macropores having an interpore distance of 1.5 µm. The macropores have diameters of ~800 nm, depths of 1.0 µm and hemispherical pore bottoms (Fig. S2). We dropped 1 µL of a solution of ATM in DMF onto the hydrophilized macroporous PDMS molds (Fig. 1a). Then, we covered the hydrophilized macroporous PDMS molds with hydrophilized SiO$_2$-covered Si wafer pieces (Fig. 1b). After a wait time of 2 h, the DMF had completely evaporated and we gently detached the macroporous PDMS molds. As a result, we obtained SiO$_2$-covered Si wafers patterned with regular arrays of ATM microparticles (Fig. 1c) extending several cm$^2$ (for a large-area SEM image see Fig. 2a). As revealed by the analysis of 44 ATM microparticles in the SEM image displayed in Fig. 2b, the ATM microparticles had a mean apparent area of 0.36 µm$^2$ ± 0.05 µm$^2$. Their height, as revealed by AFM, amounted to ~160 nm.

A previously reported route to MoS$_2$ involves at first the conversion of ATM into MoS$_3$ in the temperature range from 120°C to 360 °C followed by conversion of MoS$_3$ to MoS$_2$ in an inert atmosphere at temperatures above 800 °C [30, 33]. However, direct annealing of ATM in an inert gas atmosphere even at temperatures as high as 1000 °C does not yield high-quality MoS$_2$, while in hydrogen atmosphere the conversion of ATM to MoS$_2$ already takes place at ~ 425 °C following Eq. 1 [30].

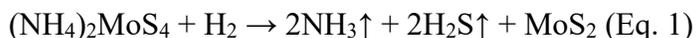

$$(NH_4)_2MoS_4 + H_2 \rightarrow 2NH_3\uparrow + 2H_2S\uparrow + MoS_2 \quad (Eq.\ 1)$$

Here, we applied a conversion following Eq. 1 and annealed the ATM microparticle arrays in the presence of sulfur under Ar/H$_2$ flow at 700 °C for 60 min (Fig. 1c, d). In this way, we obtained regular large-area arrays of MoS$_2$ microparticles (Fig. 3a, S6). The nearest neighbor distances (Fig. S7g, h) amounted to 1.38 µm ± 0.05 µm for the ATM microparticle arrays, and to 1.34 µm ± 0.10 µm for the MoS$_2$ microparticle arrays.

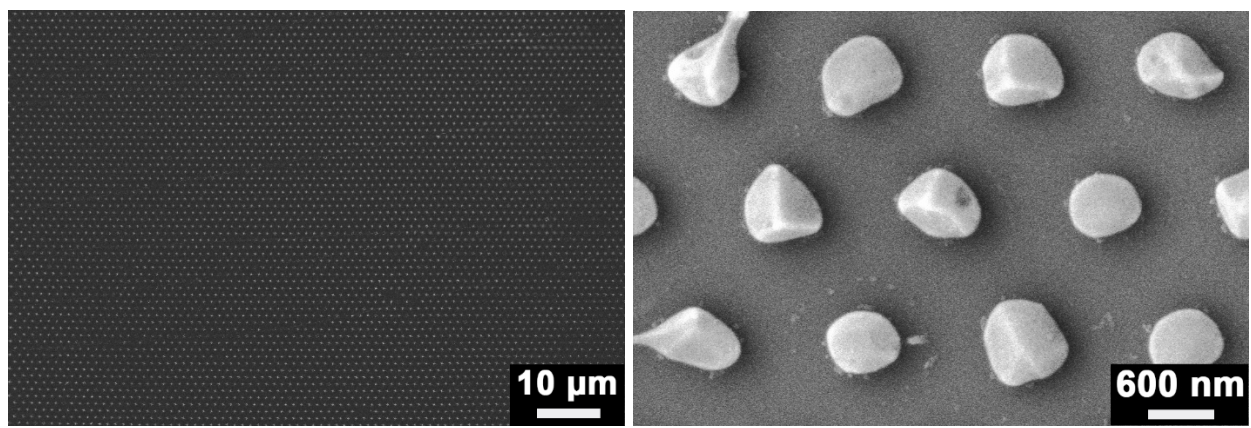

a) b)

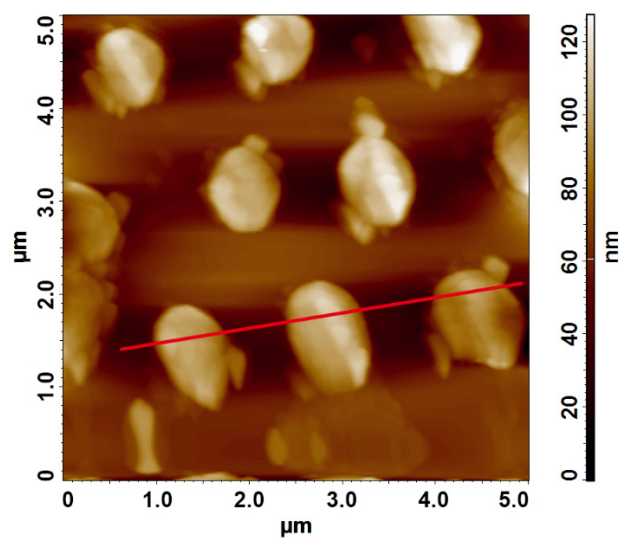
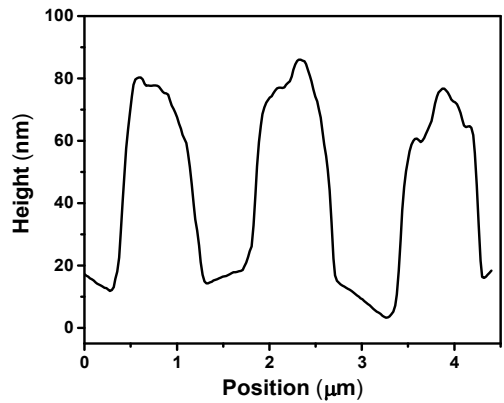

c) d)

**Fig. 3.** MoS$_2$ microparticle arrays on SiO$_2$-coated Si wafers. a) Large-area and b) higher-magnification SEM images. c) AFM topography image and d) topographic profile along the red line in panel c).

To further evaluate the shape evolution during the conversion of ATM microparticles to MoS$_2$ microparticles, we analyzed the corresponding SEM images (ATM miocroparticles: Fig. 2b; MoS$_2$ microparticles: Fig. S5) and determined the geometric descriptors apparent area, aspect ratio and roundness (Fig. S7a-f). The mean apparent area was reduced from 0.36 µm$^2$ ± 0.05 µm$^2$ for the ATM microparticles to 0.29 µm$^2$ ± 0.07 µm$^2$ for the MoS$_2$ microparticles. The aspect ratio is the ratio of width and height of an object. The mean aspect ratio increased from 1.20 ± 0.14 for the ATM microparticles to 1.33 ± 0.28 for the MoS$_2$ microparticles. The roundness 4×area/(π×major_axis$^2$) describes how closely the contour of an object resembles that of a perfect circle having a roundness of 1. The mean roundness decreased from 0.85 ± 0.09 for the ATM microparticles to 0.78 ± 0.14 for the MoS$_2$ microparticles. Notably, the standard deviations of the apparent area, aspect ratio, roundness, and nearest neighbor distances increase after the high-temperature treatment that converted the ATM microparticles to MoS$_2$ microparticles. The decrease in the average apparent area and the mean roundness, as well as the increase in the aspect ratios indicate that shape reconstructions occurred so that the MoS$_2$ microparticles are stronger faceted than the ATM microparticles. This notion is corroborated by the detailed view of MoS$_2$ microparticles shown in Fig. 3b. Energy-dispersive X-ray (EDX) mappings revealed homogeneous distributions of Mo and S atoms within the prepared MoS$_2$ microparticles (Fig. S8). The height of the MoS$_2$ microparticles amounted to ~70 nm (Fig. 3c, d), about half the height of the ATM microparticles.

To characterize the crystal phase of the obtained MoS$_2$ microparticles, X-ray powder diffraction (PXRD) (Fig. 4a) and Raman spectroscopy (Fig. 4b) were employed. The PXRD pattern of a powder of MoS$_2$ microparticles scraped off from SiO$_2$-covered Si substrates shows diffraction peaks at 2Θ values of 14.2°, 32.7°, 39.6°, 44.1°, 49.4°, 58.5°, and 60.3°, which could be assigned to the (002), (100), (103), (006), (105), (110), and (008) reflections of hexagonal MoS$_2$ (space group P 63/mmc; JCPDF card 37-1492) in line with previous reports [34-36]. Raman spectra were measured with a confocal Raman imaging system; the diameter of the laser spot on the sample surface amounted to ~3 µm so that approximately 3 MoS$_2$ microparticles were covered (Fig. 4b). The Raman spectra acquired in this way exhibited two pronounced peaks at 383.2 cm$^{-1}$ and 407.5 cm$^{-1}$ representing the in-plane E$^1_{2g}$ and the out-of-plane A$_{1g}$ vibrational modes of MoS$_2$ [37, 38]. The peak positions and the frequency difference of 24.3 cm$^{-1}$ between the peaks indicate bulk-like properties which might be beneficial for the mechanical performance [39, 40]. A Raman mapping performed for the E$^1_{2g}$ mode of MoS$_2$ at 383.2 cm$^{-1}$ (Fig. 4c) confirmed the conversion of the ATM microparticles to MoS$_2$ microparticles under conversation of the array ordering.

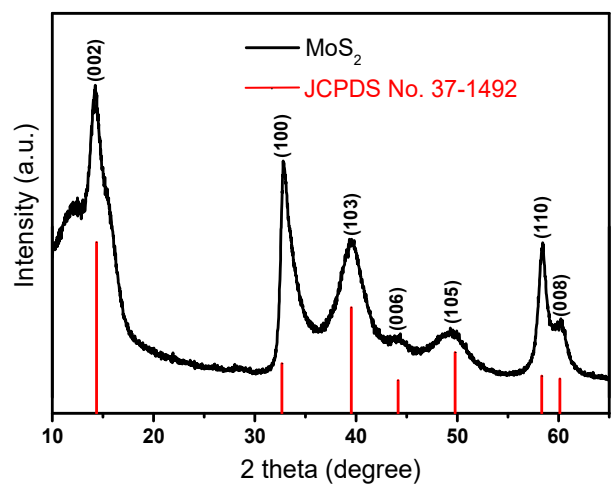

a)

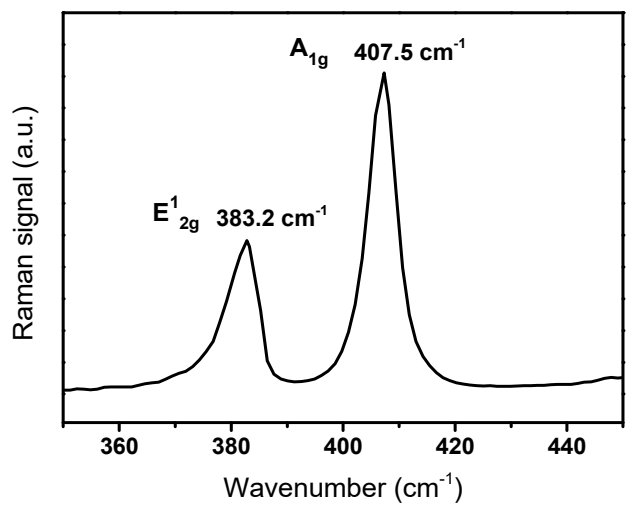

b)

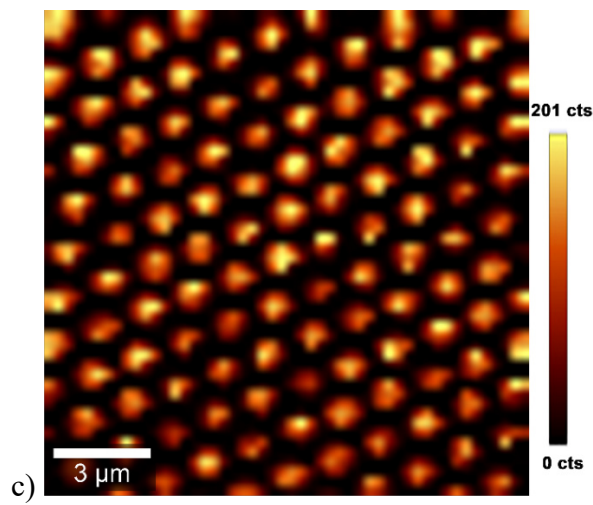

c)

**Fig. 4.** a) X-ray powder diffraction (PXRD) pattern of MoS$_2$ microparticles, b) Raman spectrum of MoS$_2$ microparticles located on a SiO$_2$-coated Si wafer and c) Raman mapping at 383.2 cm$^{-1}$ of an array of MoS$_2$ microparticles located on a SiO$_2$-coated Si wafer including 225 spectra captured in an area of 15 × 15 μm$^2$.

To transfer the MoS$_2$ microparticle arrays to the contact surfaces of PMMA monoliths, we spin-coated solutions of PMMA in chloroform onto MoS$_2$ microparticle arrays located on SiO$_2$-coated Si wafers. After the evaporation of the chloroform, a ~115 μm thick PMMA layer was obtained (Fig. S4). Partial etching of the SiO$_2$ layer in aqueous KOH solution allowed the detachment of the obtained PMMA_MoS$_2$ monoliths. The ordering of the MoS$_2$ microparticle arrays was conserved as revealed by SEM investigations on PMMA_MoS$_2$ monoliths (Fig. 5); the ordering of the MoS$_2$ microparticle arrays at the contact surfaces of the PMMA_MoS$_2$ monoliths corresponded to that of the MoS$_2$ microparticle arrays on the SiO$_2$-covered Si wafers. The corresponding AFM image (Fig. 5c) shows that the MoS$_2$ microparticles were partially embedded into the contact surface of the PMMA_MoS$_2$ monoliths. The MoS$_2$ microparticles protruded ~ 14 nm from the contact surface (Fig. 5d).

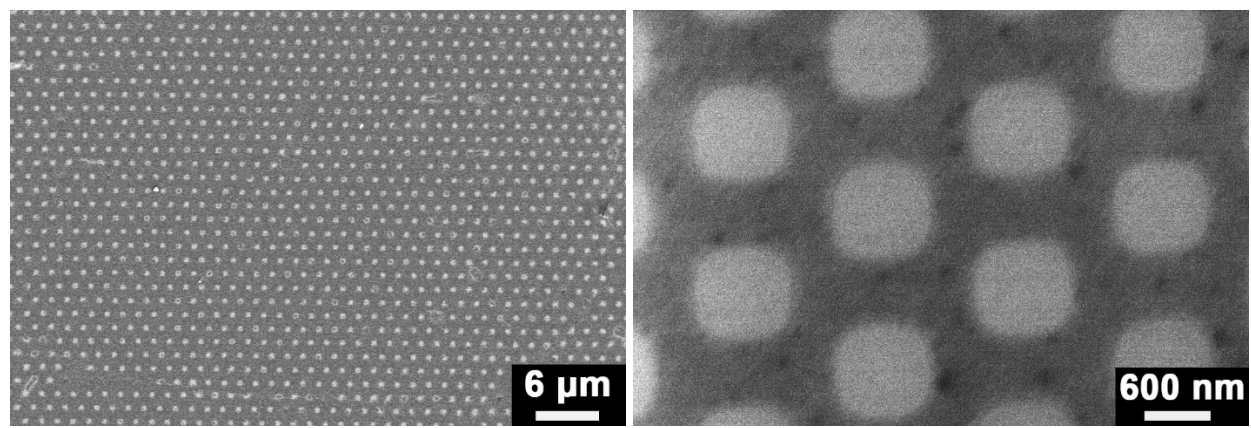

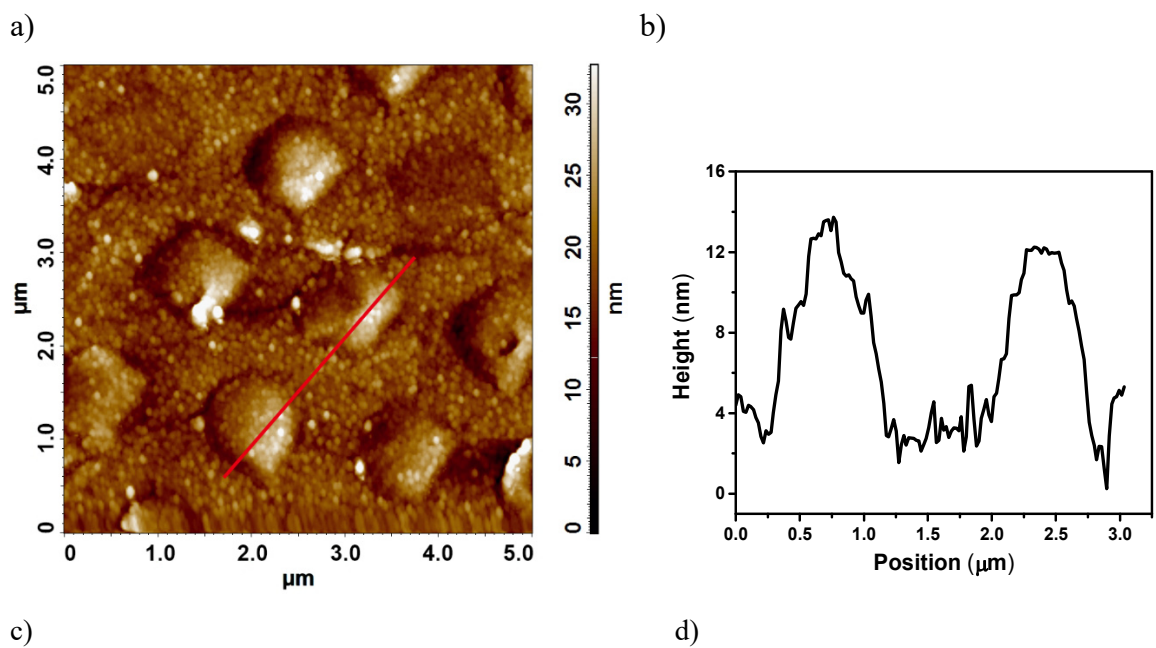

**Fig. 5.** PMMA_MoS$_2$ monoliths. a), b) SEM images, c) AFM image, d) topographic AFM line profile along the red line in panel c).

To evaluate the impact of the MoS$_2$ microparticle arrays on the interfacial tribological properties of PMMA_MoS$_2$ monoliths, we measured the torque of the PMMA_MoS$_2$ contact surfaces and – as a reference – of the contact surfaces of a PMMA monolith. The latter was prepared in the same way than the MoS$_2$_PMMA monoliths, but in the absence of MoS$_2$ microparticles. The tribological measurements were carried out using a rotating-disk configuration. We applied various normal forces resulting in contact pressures of 1929 Pa, 3858 Pa and 5787 Pa. The disk, on which the sample to be investigated had been mounted, was rotated at an angular frequency of $2.5\pi$ rad s$^{-1}$. Each sample was subjected to seven successive measurements with a duration of 17 s to determine the magnitude $\tau$ of the torque $\boldsymbol{\tau}$, from which the friction coefficients $\mu$ were calculated. In the rotating disk configuration used here, the contact area between the rotating samples and the fixed aluminum disk serving as counter surface has a radius $L = 9$ mm. Each annular ring segment of the contact area with a width d$r$ and a distance $r$ from the center ($0 \leq r \leq L$) contributes a differential torque d$\tau(r)$ to $\tau$, which depends on $r$. Hence, $\tau$ is obtained by integrating d$\tau(r)$ from $r = 0$ to $r = L$:

$$\tau = \int_0^L d\tau(r) \text{ (Eq. 2)}$$

The area of an annular ring segment amounts to $2\pi r dr$. The mass $m$ applied to the contact area represents the applied normal force and the applied contact pressure. The fraction d$m(r)$ of $m$ that can formally be designated to an annulus amounts to:

$$dm = \frac{m}{\pi L^2} 2\pi r dr = \frac{2mr}{L^2} dr \text{ (Eq. 3)}$$

The fraction d$f_{\text{friction}}$ of the friction force $f_{\text{friction}}$ that can be assigned to an annulus amounts to:

$$f_{\text{friction}} = \mu \, (dm)g = \frac{2\mu mgr}{L^2} dr \text{ (Eq. 4),}$$

where $g$ is the acceleration of gravity. The differential torque d$\tau(r)$ equals:

$$d\tau(r) = r \, df_{\text{friction}}(r) = \frac{2\mu mgr^2}{L^2} dr \text{ (Eq. 5).}$$

Combining Eq. 2 and Eq. 5 yields:

$$\tau = \int_0^L d\tau(r) = \int_0^L \frac{2\mu mgr^2}{L^2} dr = \frac{2}{3}\mu mgL \text{ (Eq. 6)}$$

Thus, the friction coefficients $\mu$ can be calculated as:

$$\mu = = \frac{3\tau}{2mgL} \text{ (Eq. 7)}$$

The average torque $\tau$ obtained by seven successive measurements on PMMA_MoS$_2$ monoliths amounted to $0.57246 \pm 0.071 * 10^{-3}$ Nm for a contact pressure of 1929 Pa, to $1.3406 \pm 0.0758 * 10^{-3}$ Nm for a contact pressure of 3858 Pa and to $1.53522 \pm 0.0467 * 10^{-3}$ Nm for a contact pressure of 5787 Pa (Fig. 6a and Supporting Table S1). The average friction coefficients for the MoS$_2$_PMMA monoliths $\mu_{\text{MoS2\_PMMA}}$ amounted to $0.19463 \pm 0.0214$ for a contact pressure of 1929 Pa, to $0.17379 \pm 0.00889$ for a contact pressure of 3858 Pa and to $0.16887 \pm 0.00514$ for a contact pressure of 5787 Pa (Fig. 6b and Table S2). Derazkola and Simchi showed that PMMA containing 5-20 vol% Al$_2$O$_3$ nanoparticles having an average diameter of 50 nm exhibits a friction coefficient of 0.2~0.5 [41]. Lin et al. reported a friction coefficient of ~0.25 for PMMA monoliths coated with bilayered films composed of diamond-like carbon and silicon [42]. Li reported a decrease in the friction coefficient from 0.59 for PMMA to 0.35 for PMMA containing 7 vol% TiO$_2$ nanoparticles [43]. Considering these benchmarks, embedding ordered MoS$_2$ microparticle arrays into the surface of PMMA monoliths appears to be an effective method to alleviate interfacial friction.

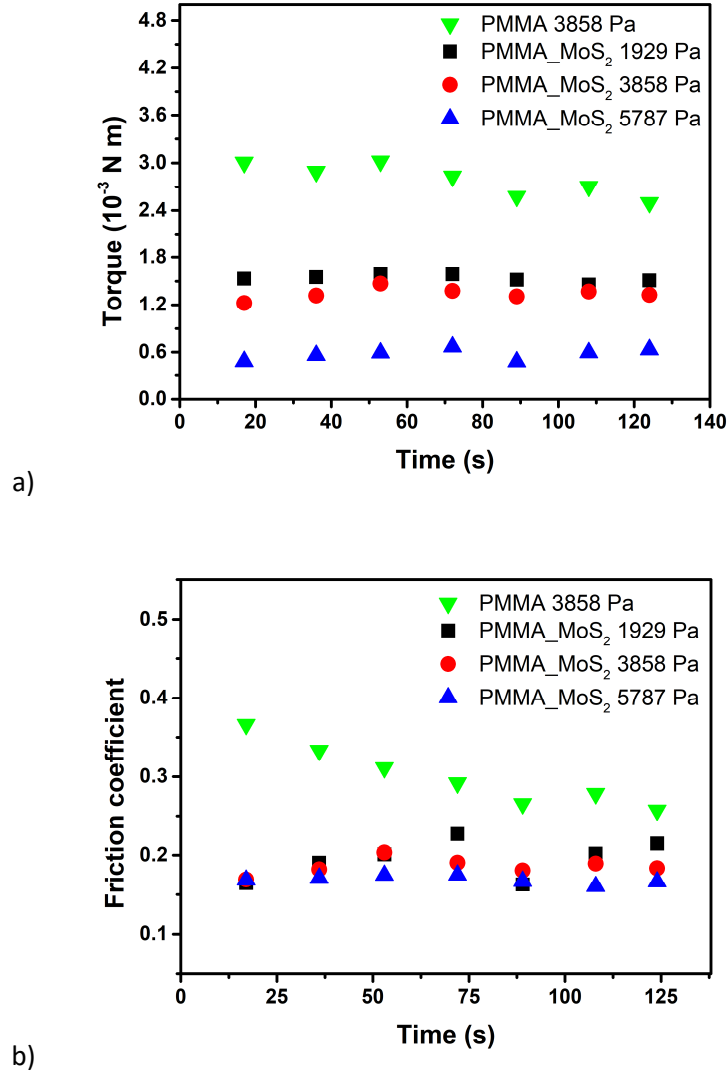

a)

b)

**Fig. 6.** a) Torque $\tau$ and b) friction coefficient $\mu$ of PMMA_MoS$_2$ monoliths and a PMMA monolith (green down-triangles) prepared in the same way than the PMMA_MoS$_2$ monoliths but in the absence of MoS$_2$ microparticles. The same PMMA_MoS$_2$ sample was successively subjected to the measurements at pressures of 1929 Pa (black squares) and 5787 Pa (blue up-triangles). The measurement at 3858 Pa was carried out on a different PMMA_MoS$_2$ sample (red circles). For each applied pressure seven successive measurements were carried out with an angular frequency of 2.5π rad s$^{-1}$ and a duration of 17s.

The surface topography of PMMA_MoS$_2$ monoliths after the seven successive torque measurements summarized in Fig. 6a was studied by SEM (Fig. 7a, for a large-scale image see

Fig. S9) and AFM (Fig. 7b and c). The large-scale ordering of the hexagonal MoS$_2$ microparticle arrays was conserved. On the local scale, the MoS$_2$ microparticles remained fixed at their initial positions. No indications of additional displacement or even detachment of MoS$_2$ microparticles were apparent apart from defects originating from preparation steps prior to the torque measurements, as revealed by a comparison of corresponding SEM images (cf. Fig. 3a). The spike-like structures seen in Figure 7a are a signature of the markedly faceted shape of the MoS$_2$ microparticles (cf. Figures 3, 4c, 5b and Supporting Figure S5) rather than a result of the torque measurements. However, we assume that the initial contact formation with the aluminum disk used for the torque measurements resulted in additional flattening of the protruding parts of the MoS$_2$ microparticles. The initial contact is formed at asperities, resulting in small actual contact areas, high compressive stress and, therefore, likely in local plastic deformation. This assumption is in line with observations reported by Tedstone et al. [44], who exerted normal compressive load on MoS$_2$ specimens with dimensions similar to those of the MoS$_2$ microparticles studied by us.

It is reasonable to assume that the torque measurements summarized in Fig. 6 were carried out in the boundary lubrication regime, where load is primarily supported by contacting protruding asperities. On the macroscopic scale, the MoS$_2$ microparticles at the surfaces of the PMMA_MoS$_2$ monoliths had no preferential crystal orientation with respect to the directions of normal load or of friction. However, already Martin et al. reported friction-induced orientation of the MoS$_2$ basal planes parallel to the sliding direction [45]. Tedstone et al., whose MoS$_2$ specimens were also characterized by the absence of preferential crystal orientations with respect to external reference directions, assumed that application of compressive forces aligns the basal crystallographic planes of MoS$_2$ with the substrate [44]. AFM images of unoriented MoS$_2$ surfaces are typically characterized by isotropic grain structures [45]. However, the AFM images we obtained from our samples after the torque measurements (cf. Figure 7b) show signs of orientation, which can possibly be interpreted as the result of the impact of shear. Therefore, we speculate that load-induced and shear-induced orientation of the MoS$_2$ basal planes possibly contributed to the observed experimental outcome. The moderate decrease in the torque values measured on PMMA_MoS$_2$ monoliths with increasing contact pressure could be an indication for more pronounced orientation of the MoS$_2$ basal planes parallel to the monolith surface.

The average torque measured for a PMMA monolith used as reference sample at a contact pressure of 3858 Pa amounted to $2.79173 \pm 0.2039 * 10^{-3}$ Nm, and the corresponding friction coefficient to 0.35946 ± 0.01992. Both values are twice as high as the values obtained on a PMMA_MoS$_2$ monolith tested under the same contact pressure. A direct systematic comparison of the friction properties of both samples is not possible because the real contact areas are likely different. However, from a pragmatic application-related point of view, the incorporation of the MoS$_2$ microparticle arrays into the surfaces of the PMMA_MoS$_2$ monoliths results in an enhanced sliding capability. It should be noted that torque and friction coefficient of the tested PMMA monolith decreased in the course of successive measurements. This outcome is possibly related to shear-induced cleavage of PMMA chains at the surface of the PMMA monolith. If so, the functionalization of the surfaces of PMMA monoliths with MoS$_2$ microparticles would not only increase the sliding capability but would also protect the monolith itself from shear-induced damage.

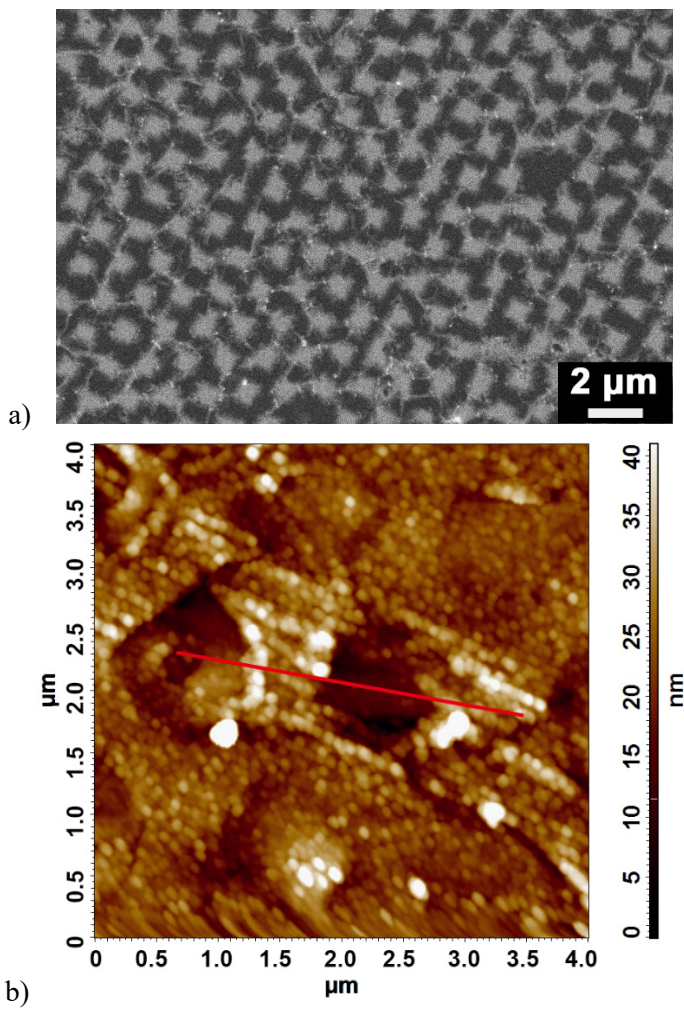

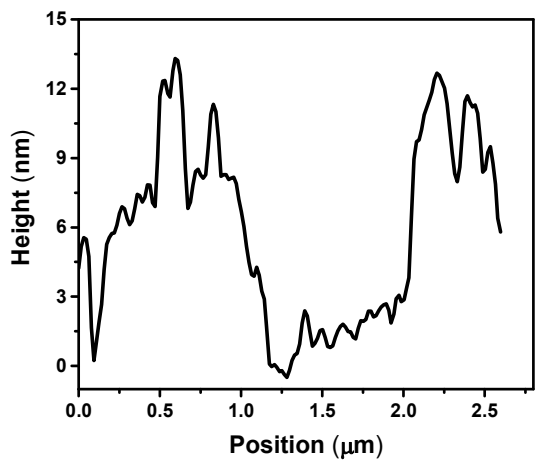

c)

**Fig. 7.** a) SEM image and b) AFM image of a PMMA_MoS$_2$ monolith after seven successive torque measurements; d) topographic AFM line profile along the red line in panel c).

## 4. Conclusions

Parts and components of devices and other types of technical systems often need to be self-lubricating. Therefore, the tribological properties of their contact surfaces need to be tailored, typically by means of dry lubricants. Continuous layers of dry lubricants often suffer from insufficient adhesion between dry lubricant and substrate, delamination and crack propagation. We have evaluated whether these drawbacks can be overcome by replacing continuous coatings with dense ordered arrays of dry lubricant microparticles embedded into the substrate surface. It is obvious that delamination and crack propagation are intrinsically impossible if substrate surfaces are modified with microparticle arrays. Moreover, the partial embedding of the microparticles into the substrate surfaces way may enhance adhesion between dry lubricant and the underlying bulk substrate. We selected the widely used dry lubricant MoS$_2$ as model system. A PDMS mold with macropore arrays fabricated by double replication of macroporous silicon templates was employed to prepare monodomains of ATM microparticles on SiO$_2$-coated Si wafers. Subsequent high-temperature conversion under Ar/H$_2$ flow in the presence of elemental sulfur yielded large-area hexagonal arrays of MoS$_2$ microparticles with diameters of ~800 nm and heights of ~70 nm. The MoS$_2$ microparticle arrays were transferred to the surfaces of PMMA monoliths under conservation of the array ordering. Tribological tests revealed that the presence of the MoS$_2$ microparticle arrays at the surfaces of PMMA monoliths halves the friction coefficient. The results reported here indicate that tailoring of the tribological properties of parts and components consisting of polymers and other moldable materials can be achieved by embedding arrays of microparticles into their surfaces, thus circumventing the problems with continuous coatings mentioned above.

## CRediT authorship contribution statement

**Weijia Han:** Investigation, Visualization, Methodology, Writing. **Siwei Luo:** Investigation, Formal analysis. **Dirk Bröker:** Investigation. **Norbert Vennemann:** Investigation, Formal analysis. **Markus Haase:** Investigation. **Georg S. Duesberg:** Investigation, Resources, Funding acquisition. **Martin Steinhart:** Conceptualization, Supervision, Funding acquisition, Project administration, Resources, Writing - review & editing.

## Declaration of Competing Interest

The authors declare no conflict of interest.

## Acknowledgements

The authors thank the European Research Council (ERC-CoG-2014, Project 646742 INCANA) for funding. GSD thanks the European Commission under the project Graphene Flagship (881603) for financial support. Support with PXRD measurements by Dr. C. Homann is gratefully acknowledged.

## Supporting information

Supporting information is available from the https://www.sciencedirect.com or from the author.